\documentclass[journal, one column]{IEEEtran}
\usepackage{amsmath, amsthm, amsfonts}
\usepackage[english]{babel}
\usepackage[dvips]{graphicx}
\usepackage{graphicx}
\usepackage{tabularx}
\usepackage{enumerate}
\usepackage{mathtools}

\usepackage{setspace}
\begin{document}
\title{Adaptive Convex Combination of APA and ZA-APA algorithms for Sparse System Identification}


\author{{Vinay Chakravarthi Gogineni$^1$, Mrityunjoy Chakraborty$^2$, Senior Member, IEEE}\\
Department of Electronics and Electrical Communication Engineering\\
Indian Institute of Technology, Kharagpur, INDIA\\
Phone: $+91-3222-283512 \hspace{2em}$ Fax: $+91-3222-255303$\\
E.Mail : $^1\;$vinaychakravarthi@ece.iitkgp.ernet.in, $^2\;$mrityun@ece.iitkgp.ernet.in}

\maketitle
\thispagestyle{empty}

\begin{abstract}
In general, one often encounters the systems that have sparse impulse response,
with time varying system sparsity. Conventional adaptive filters which
perform well for identification of non-sparse systems fail to exploit
the system sparsity for improving the performance as the sparsity level
increases. This paper presents a new approach that uses an adaptive
convex combination of Affine Projection Algorithm (APA) and
Zero-attracting Affine Projection Algorithm (ZA-APA)algorithms
for identifying the sparse systems, which adapts dynamically to the
sparsity of the system. Thus works well in both sparse and non-sparse
environments and also the usage of affine projection makes it robust
against colored input. It is shown that, for non-sparse systems, the
proposed combination always converges to the APA algorithm, while for
semi-sparse systems, it converges to a solution that produces lesser
steady state EMSE than produced by either of the component
filters. For highly sparse systems, depending on the value of the
proportionality constant ($\rho$) in ZA-APA algorithm, the proposed
combined filter may either converge to the ZA-APA based filter or
produce a solution similar to the semi-sparse case i.e., outerperforms
both the constituent filters.
\end{abstract}

\section{Introduction}
Usually, many real-life systems exhibit sparse representation
i.e., their system impulse response is characterized by small
number of non zero taps in the presence of large number of
inactive taps. Sparse systems are encountered in many important
practical applications such as network and acoustic echo
cancelers $\cite{1}$-$\cite{2}$, HDTV channels $\cite{3}$,
wireless multipath channels $\cite{4}$, underwater acoustic
communications $\cite{5}$. The conventional
system identification algorithms such as LMS and NLMS
are sparsity agnostic i.e., they are unaware of
underlying sparsity of the system impulse response.  Recent
studies have shown that the \emph{a priori} knowledge about the
system sparsity, if utilized properly by the identification
algorithm, can result in substantial improvement in its estimation
performance. This resulted in a flurry of research activities in
the last decade or so towards developing sparsity aware adaptive
filter algorithms, notable amongst them being the Proportionate
Normalized LMS (PNLMS) algorithm $\cite{6}$ and its variants
$\cite{7}$-$\cite{8}$. These current regressor based algorithms exhibit
slower convergence rate for the correlated input. As a solution, these proportionate-type
concepts were extended to data reuse case to provide the uniform convergence rate for both white
and correlated input $\cite{9}$.
\par
On the other hand, drawing from the ideas of compressed Sensing (CS)
$\cite{10}$-$\cite{12}$, several new sparse adaptive filters have
been proposed in the last few years, notably, the zero attracting LMS
(ZALMS) $\cite{13}$ derived by incorporating the $l_{1}$ norm penalty
into the LMS cost function which was later extended to ZA-APA $\cite{14}$. The simplicity of using
$l_{1}$ norm penalty makes their implementation extremely simple. These algorithms perform well
for the systems that are highly sparse, but struggle as the
system sparsity decreases. That means these algorithms cannot
perform well as the system sparsity varies widely over time.
\par
In $\cite{15}$, an improved PNLMS (IPNLMS)
algorithm, which is a controlled mixture of the
PNLMS and the NLMS is proposed to handle the variable
system sparsity. In $\cite{16}$, an adaptive convex
combination of two IPNLMS filters is proposed that can
adapt to situations where the system sparsity is
time varying and unknown. However, it provides the steady
state MSD which is same as that of conventional sparse agnostic filters.
Reweighted ZALMS (RZALMS) $\cite{13}$ addresses this issue by selecting
the shrinkage parameter $\rho$, in a tricky fashion at the
cost of increased complexity. As a solution to this problem,
in $\cite{17}$, a convex combination of LMS and ZALMS algorithm
has been proposed that switches between ZALMS and LMS adaptive
filters depending on the sparsity level, thus enjoying the robustness
against time-varying system sparsity with reduced complexity compared
to the RZALMS. However, the aforementioned mechanisms that support time varying sparsity could
not perform well in colored (correlated) input signal condition.
\par
In this paper, we propose an alternative method that enjoys the robustness
against time-varying system sparseness as well as coloredness of
(correlated) input signal by using an adaptive convex combination
of the Affine Projection Algorithm (APA) and Zero Attracting Affine
Projection Algorithm (ZA-APA). The proposed algorithm uses the
general framework of convex combination of adaptive filters
and requires less complexity. The performance of the proposed
scheme is first evaluated analytically by carrying out the convergence
analysis. This requires evaluation of the steady state cross correlation
between the output \emph{a priori} errors generated by the APA and the
ZA-APA based filters and then relating it to the respective steady
state EMSE of the two filters. In our analysis, we have carried
out this exercise for three different kinds of systems, namely, non-sparse,
semi-sparse and sparse. The analysis shows that the proposed combined
filter always converges to the APA based filter for a non-sparse
system, while for semi-sparse systems, it converges to a solution
that can outperform both the constituents. For sparse systems,
the proposed scheme usually converges to the ZA-APA based filter.
However, by adjusting the regularized parameter $\rho$, a solution
like the semi sparse case can be achieved i.e., outperforming
both the filters. Finally, the robustness of the proposed methods
against variable sparsity and coloredness is verified by the
detailed simulation studies.
\section{Problem Formulation, Proposed Algorithm and Performance Analysis}
We consider the problem of identifying an unknown system (supposed to
be sparse), modeled by the $L$ tap coefficient vector $\textbf{w}_{opt}$
which takes a signal $u(n)$ with variance $\sigma_{u}^{2}$
as the input and produces the observable output
$d(n)=\textbf{u}^{T}(n) \textbf{w}_{opt} + {\epsilon}(n)$, where
$\textbf{u}(n)=[ {u}(n), {u}(n-1), . . . , {u}(n-M+1)]^{T}$ is the
input data vector at time $n$, and $\epsilon(n)$ is the observation
noise with zero mean and variance $\xi^{0}$. In order to identify the
system, the approach mentioned in $\cite{18}$ is followed to deploy an
adaptive convex combination of two adaptive filters as
shown in the Fig. $1$, where $l^{th}$ filter uses the APA/ZA-APA algorithm
to adapt a filter coefficient vector $\textbf{w}_{l}(n)$ as follows,
\begin{equation} \label{eq1}
\begin{split}
{\textbf{w}}_{l}(n+1)={\textbf{w}}_{l}(n)+ \mu \mspace{4mu} {\textbf{U}}(n) \Big( \varepsilon \textbf{I}_{M} +{\textbf{U}}^{T}(n) {\textbf{U}}(n) \Big)^{-1} {\textbf{e}}_{l}(n) - \rho_{l} \mspace{4mu} sgn\left({\textbf{w}}_{l}(n)\right)
\end{split}
\end{equation}
where $\mu$ is the step size (common for both the filters), controller constant $\rho_{l}=0$,
when $l=1$ and $\rho_{l}= \rho$, for $l=2$, also $\textbf{e}_{l}(n) = \textbf{d}(n)-
{\textbf{y}}_{l}(n)= [e_{l}(n), e_{l}(n-1), . . . , e_{l}(n-M+1)]^{T}$ is
the respective filter output error vectors with $\textbf{y}_{l}(n)=
\textbf{U}^{T}(n) \textbf{w}_{l}(n)$ denoting the respective filter
output vectors, and ${\textbf{U}}(n)= [\textbf{u}(n), \textbf{u}(n-1),
. . . , \textbf{u}(n-M+1)]^{T}$ is the input data matrix. The convex
combination generates a combined output vector $y(n)=\lambda(n)
\mspace{2mu}y_{1}(n) +[1-\lambda(n)]\mspace{2mu}y_{2}(n)$.
The variable $\lambda(n) \in [0, 1]$, is a mixing parameter,
which is to be adapted by the following gradient descent method to minimize the
quadratic error function of the overall filter, namely $e^{2}(n)$, where $e(n)=
d(n)-y(n)$. However, such adaptation does not guarantee that $\lambda(n)$ will
lie between 0 and 1. Therefore, instead of $\lambda(n)$, an equivalent variable
$a(n)$ is updated which expresses $\lambda(n)$ as a sigmoidal function, i.e.,
$\lambda(n)=\frac{1}{1+ exp  \left(-a(n)\right)}$. The update equation of
$a(n)$ is given by $\cite{18}$,
\begin{equation}\label{eq2}
\begin{split}
a(n+1)=a(n) - \frac{\mu}{2} \mspace{4mu} \frac{\partial{e^{2}(n)}}{\partial{a(n)}}
=a(n) + \mspace{4mu} \mu_{a} \mspace{4mu} e(n) \mspace{4mu} [y_{1}(n) - y_{2}(n)] \lambda(n) [1- \lambda(n)]
\end{split}
\end{equation}
In practice, $\lambda(n) \approx 1$ for $a(n) \gg 0$ and conversely,
$\lambda(n) \approx 0$ for $a(n) \ll 0$. Therefore, instead of updating
the $a(n)$ up to $\pm \infty$, it is sufficient to restrict it to a
range $[-a^{+}, +a^{+}]$ ($a^{+}:$ a large finite number) which limits
the permissible range of $\lambda(n)$ to $[1-\lambda^{+}, \lambda^{+}]$,
where $\lambda^{+}= \frac{1}{1+ exp(-a^{+})}$.
\begin{figure} [h]
\centering
\includegraphics [height=70mm,width=110mm]{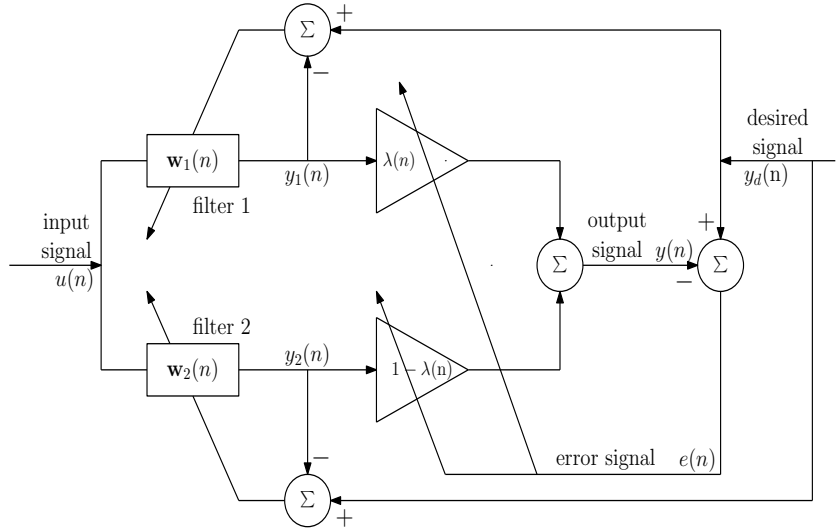}
\caption{The proposed adaptive convex combination of two adaptive filters ( Filter $1$ : ZA-APA  and Filter $2$ : APA}
\label{the-label-for-cross-referencing}
\end{figure}
\subsection{Important Assumptions and Definitions}
In $\cite{19}$, $\cite{20}$ convergence analysis of APA was presented using an
equivalent form of APA known as Normalized LMS with Orthogonal Correction Factors
(NLMS-OCF), along with the assumptions proposed by Slock in $\cite{21}$ for
modeling the system input. The analysis in $\cite{19}$ provides
the basis for the work presented in this paper. Some introduction
about the NLMS-OCF algorithm is presented below. For the given input signal $u(n)$,
observation noise $\epsilon (n)$ and the reference signal $d (n)$,
the NLMS-OCF algorithm uses orthogonal correction factors based on the
past $M-1$ input vectors (each of length $L$), to update the weights
$\textbf{w}(n)$, in every iteration. The weight update equation is
\begin{equation}\label{eq3}
\begin{split}
\textbf{w}(n+1)&= \textbf{w}(n)+ \mu_{0} \hspace{0.3em} \textbf{u}(n)  + \mu_{1} \hspace{0.3em} \textbf{u}^{1}(n)
+ . . . + \mu_{M-1} \hspace{0.3em} \textbf{u}^{M-1}(n)
\end{split}
\end{equation}
where the notation $\textbf{u}^{k}(n)$ refers to the
component of vector $\textbf{u}(n-k)$ that is orthogonal to the
vectors: $\textbf{u}(n), \textbf{u}(n-1), . . . , \textbf{u}(n-k+1)$,
where $k=0, 1, . . . , M-1$ is the previous input signal vector index.\par
And also $\mu_{k}$ for $k = 1, 2, \hdots, M-1$ is chosen as,
\begin{equation}\label{eq4}
\begin{split}
\mu_{k}=
\begin{cases}
\frac{\mu e(n)}{\textbf{u}^{T}(n)\textbf{u}(n)},  \hspace{6.4em} \text{for} \hspace{1em} k=0, \hspace{1em}  \text{if} \hspace{1em} \|\textbf{u}(n)\|\neq 0\\
\frac{\mu e^{k}(n)}{(\textbf{u}^{k}(n))^{T}\textbf{u}^{k}(n)},  \hspace{5em} \text{for} \hspace{1em} k=1, 2,\hdots, M-1, \hspace{1em}  \text{if} \hspace{1em} \|\textbf{u}^{k}(n)\|\neq 0
\end{cases}
\end{split}
\end{equation}
where
\begin{equation}\label{eq5}
\begin{split}
e(n)&= d(n)-\textbf{w}^{T}(n) \textbf{u}(n)\\
e^{k}(n)&= d(n-k)-(\textbf{w}^{k}(n))^{T}\textbf{u}(n-k) \mspace{20mu}    \textit{for} \mspace{8mu} k= 1, 2, \hdots, M-1 \\
\textbf{w}^{k}(n)&= \textbf{w}(n)+ \mu_{0} \hspace{0.3em} \textbf{u}(n)  + \mu_{1} \hspace{0.3em} \textbf{u}^{1}(n)
+ . . . + \mu_{k-1} \hspace{0.3em} \textbf{u}^{k-1}(n)
\end{split}
\end{equation}
The algorithm can be seen as the process of computing
weight updates, using NLMS, based on the current
input data vector, $\textbf{u}(n)$, as well as the
orthogonal components from each of the previous $M-1$
input data vectors.
\par
In addition to NLMS-OCF, as described in $\cite{19}$ the performance analysis is done based
on the following assumptions on the signal and the underlying system
\begin{enumerate}
\item [A1)]The input signal vectors $\textbf{u}(n)$ are assumed to be zero mean with covariance matrix
  \begin{equation}\label{eq6}
   \textbf{R}=E[\textbf{u}(n) \textbf{u}^{T}(n)]={\textbf{V}}{\bf{\Lambda}}{\textbf{V}}^{T}
  \end{equation}
   where ${\bf{\Lambda}}= diag( \lambda_{1}, \lambda_{2}, . . . , \lambda_{L})$, and ${\textbf{V}}= ( \textbf{\emph{v}}_{1}, \textbf{\emph{v}}_{2},
 . . . , \textbf{\emph{v}}_{L})$. Here, $\lambda_{1}, \lambda_{2}, . . . , \lambda_{L}$ are the eigenvalues of input covariance matrix $\textbf{R}$ and $\textbf{\emph{v}}_{1}, \textbf{\emph{v}}_{2},
 . . . , \textbf{\emph{v}}_{L}$ are the corresponding orthonormal eigenvectors $(\textbf{V}^{T}\textbf{V}=\textbf{I})$ i.e., $\textbf{V}$ is a unitary matrix.

 \item [A2)]Observation noise $\epsilon(n)$ (i.e., zero mean white noise, with variance $\xi^{0}$) is independent of $\textbf{u}(n)$ and,  the initial conditions $w_{i}(0)$ and $a(0)$ are also independent of $\textbf{u}(n)$, $d(n)$.

\item [A3)]The random signal vector $\textbf{u}(n)$ is the product of three i. i. d. random variables, that is
  \begin{subequations}
    \begin{equation} \label{eq7a}
   \textbf{u}(n)= s(n) r(n) \textbf{\emph{v}}(n)
   \end{equation}
   \text{ where}
   \begin{equation}  \label{eq7b}
    \begin{split}
  P\{s(n)=\pm1\}&=\frac{1}{2}\\
  r(n)&\sim \|\textbf{u}(n)\| \\
 p\{\mathbf{v}(n) = \textbf{\emph{v}}_{i}\}&= p_{i}= \frac{\lambda_{i}}{tr(\mathbf{R})}, \hspace{2em} i=1, 2, . . . , L
  \end{split}
  \end{equation}
    \end{subequations}
\end{enumerate}
here $r(n)\sim \|\textbf{u}(n)\|$ i.e., $r(n)$ has the same
distribution as the norm of the true input signal vectors.
\par
Assumption A3 was first introduced by Slock in $\cite{19}$, which leads to a simple distribution for the vectors $\textbf{u}(n)$
while following the actual first and second order statistics of
the input signal in A1 consistently. Assumption A3 was used in $\cite{19}$, $\cite{20}$
as well as here, to To simplify the convergence analysis.
\par
As mentioned in $\cite{19}$, using Assumption A3, the weight update in $\eqref{eq3}$-$\eqref{eq5}$ can be simplified, since the computation of orthogonal components $\textbf{u}^{k}(n)$ becomes unnecessary i.e., each $\textbf{u}(n)$ is already chosen from the orthogonal set. Hence, NLMS-OCF update equation can be rewritten as follows
\begin{equation}\label{eq8}
\begin{split}
\textbf{w}(n+1)&= \textbf{w}(n)+ \mu_{0} \hspace{0.3em} \textbf{u}(n)  + \mu_{1} \hspace{0.3em} \textbf{u}(n-1)
+ . . . + \mu_{M-1} \hspace{0.3em} \textbf{u}(n-M+1)\\
e(n)&= d(n)-\textbf{w}^{T}(n) \textbf{u}(n),  \hspace{3em} \text{and}\\
e^{k}(n)&= d(n-k)-\textbf{w}^{T}(n)\textbf{u}(n-k) \mspace{20mu}    \textit{for} \mspace{8mu} k= 1, 2, \hdots, M-1 \\
\end{split}
\end{equation}
where
\begin{equation}\label{eq9}
\begin{split}
\mu_{k}=
\begin{cases}
\frac{\mu e(n)}{\textbf{u}^{T}(n)\textbf{u}(n)},  \hspace{6.4em} \text{for} \hspace{1em} k=0, \hspace{1em}  \text{if} \hspace{1em} \|\textbf{u}(n)\|\neq 0\\
\frac{\mu e^{k}(n)}{\textbf{u}^{T}(n-k)\textbf{u}(n-k)},  \hspace{4.2em} \text{for} \hspace{1em} k=1, 2,\hdots, M-1, \hspace{1em}  \text{if} \hspace{1em} \|\textbf{u}^{k}(n)\|\neq 0
\end{cases}
\end{split}
\end{equation}
Therefore, using the above NLMS-OCF approximation, weight update equations of Filter$1$ and Filter$2$ are as follows \newline
APA weight update Equation:
\begin{equation}  \label{eq10}
\begin{split}
\textbf{w}_{1}(n+1)= \textbf{w}_{1}(n)+ \mu_{0} \hspace{0.3em} \textbf{u}(n)  + \mu_{1} \hspace{0.3em} \textbf{u}(n-1)
+ . . . + \mu_{M-1} \hspace{0.3em} \textbf{u}(n-M+1)
\end{split}
\end{equation}
and ZA-APA weight update equation:
\begin{equation}  \label{eq11}
\begin{split}
\textbf{w}_{2}(n+1)= \textbf{w}_{2}(n)+ \mu_{0} \hspace{0.3em} \textbf{u}(n)  + \mu_{1} \hspace{0.3em} \textbf{u}(n-1)
+ . . . + \mu_{M-1} \hspace{0.3em} \textbf{u}(n-M+1)  - \rho \hspace{0.32em} sgn\big(\textbf{w}_{2}(n)\big)
\end{split}
\end{equation}
where $\mu_{k}$ can be calculated using $\eqref{eq9}$.
\par
Next, as in $\cite{18}$, certain definitions that are useful in the
analysis are presented below. For $l=1, 2$, we thus define,
\begin{enumerate}
\item [a)]\emph{Weight Error Vectors}: $\widetilde{\textbf{w}}_{l}(n)= \textbf{w}_{opt}-\textbf{w}_{l}(n)$;
\item [b)]\emph{Equivalent Weight Vector} for the combined filter: $\textbf{w}_{c}(n)= \lambda(n) \textbf{w}_{1}(n) + [1 - \lambda(n)] \textbf{w}_{2}(n)$;
\item [c)]\emph{Equivalent Weight Error Vector} for the combined filter: $\widetilde{\textbf{w}}_{c}(n)= \textbf{w}_{opt}-\textbf{w}_{c}(n)= \lambda(n) \mspace{2mu} \widetilde{\textbf{w}}_{1}(n) + [1 - \lambda(n)] \mspace{2mu} \widetilde{\textbf{w}}_{2}(n)$;
\item [d)]\emph{A Priori Errors}: $e_{a, l}(n)= \textbf{u}^{T}(n) \widetilde{\textbf{w}}_{l}(n)$ and $e_{a}(n)= \textbf{u}^{T}(n) \widetilde{\textbf{w}}_{c}(n) $. Clearly, $e_{a}(n)= \lambda(n) \mspace{2mu} e_{a, 1}(n) + [1 - \lambda(n)] \mspace{2mu} e_{a, 2}(n)$ and $e(n)= e_{a}(n)+\varepsilon(n)$;
\item [e)]\emph{Excess Mean Square Error (EMSE)}: $ J_{ex,l}(n)=E[e^{2}_{a, l}(n)]$, for $l=1, 2$ and $J_{ex}(n)=E[e^{2}_{a}(n)]$;
\item [f)]\emph{Cross EMSE}: $J_{ex,12}(n)=E[e_{a, 1}(n) e_{a, 2}(n)] \leq \sqrt{J_{ex, 1}(n)}\sqrt{J_{ex, 2}(n)}$ [from Cauchy-Schwartz inequality]. This means $J_{ex,12}(n)$ cannot be greater than both $J_{ex,1}(n)$ and $J_{ex,2}(n)$ simultaneously.
\end{enumerate}
From $\eqref{eq2}$, one can write,
\begin{equation}\label{eq12}
\begin{split}
E[a(n+1)]= E[a(n)] + \mspace{4mu} \mu_{a} \mspace{4mu} E[e(n) \mspace{4mu} (y_{1}(n) - y_{2}(n)) \lambda(n) (1- \lambda(n))]
\end{split}
\end{equation}
As shown in $\cite{18}$, the convergence of $E[a(n)]$ in $\eqref{eq12}$
depends on the steady state values of the individual filter EMSE and the
cross EMSE, namely, $J_{ex,l}(\infty)= \lim\limits_{n \to \infty}J_{ex,l}(n)$
and $J_{ex,12}(\infty)=\lim\limits_{n \to \infty} J_{ex,12}(n)$ respectively.
In practice, both $J_{ex, l}(n)$ and  $J_{ex, 12}(n)$, however, take only
finite number of steps to reach their steady state values as both the APA
and the ZA-APA algorithms converge in finite number of iterations.
Substituting $e(n)=e_{a}(n)+\varepsilon(n)$ in $\eqref{eq12}$, where
$e_{a}(n)$ is defined above, noting that $y_{1}(n)-y_{2}(n)=e_{a,2}(n)-e_{a,1}(n)$
and also that $E[\varepsilon(n)]=0$, and assuming like $\cite{18}$ that
in steady state, $\lambda(n)$ is independent of the \textit{a priori errors}
$e_{a,l}(n)$, it is easy to verify that for large $n$
(theoretically for $n \to \infty$),
\begin{equation}\label{eq13}
\begin{split}
E[a(n+1)]= E[a(n)] + \mspace{2mu} \mu_{a} \mspace{2mu} E[\lambda(n) \mspace{2mu} [1 - \lambda(n)]^{2}] \Delta J_{2} - \mspace{2mu} \mu_{a} \mspace{2mu} E[\lambda^{2}(n) \mspace{2mu} [1 - \lambda(n)] ]\Delta J_{1}
\end{split}
\end{equation}
where $\Delta J_{1} = J_{ex, 1} (\infty) - J_{ex, 12} (\infty) $
and $\Delta J_{2} = J_{ex, 2} (\infty) - J_{ex, 12} (\infty) $.
By assuming $\Delta J_{1}$ and $\Delta J_{2}$ are constant
( i.e., APA and ZA-APA algorithms have converged) Eq. $\eqref{eq13}$,
can be used to yield the dynamics of the evolution of $E(a(n)]$. To analyze
the convergence of $E[a(n)]$, we need to evaluate
$\Delta J_{1}$ and $\Delta J_{2}$ of the proposed combination.
\section{Performance Analysis of the Combination}
In this section, we examine the convergence behavior of the
proposed convex combination for various levels of sparsity i.e., how the filter coefficients vector $\textbf{w}_{c}(n)$ of the
proposed combination adapts dynamically to an optimum value
as dictated by the sparsity of the system.
For this, first we evaluate $J_{ex, 2}(\infty)$ and $J_{ex, 12}(\infty)$.
\subsection{Mean Convergence Analysis of the ZA-APA Algorithm}
From Eq.$\eqref{eq11}$, the recursion for the weight error vector of the ZA-APA algorithm can be written as follows:
\begin{equation}\label{eq14}
\begin{split}
\widetilde{\textbf{w}}_{2}(n+1)&=\bigg[\textbf{I}_{L}-\mu \sum\limits_{j\in J_{n}}^{} \frac{\textbf{u}(n-j) \textbf{u}^{T}(n-j)}{\textbf{u}^{T}(n-j)\textbf{u}(n-j)} \bigg] \widetilde{\textbf{w}}_{2}(n)
 -\mu \sum\limits_{j\in J_{n}}^{} \frac{\textbf{u}(n-j) \varepsilon(n-j)}{\textbf{u}^{T}(n-j)\textbf{u}(n-j)} \hspace{.2em}  + \rho \hspace{0.2em} sgn\big(\textbf{w}_{2}(n)\big)
\end{split}
\end{equation}
where $J_{n} \subseteq \{0, 1, \hdots , M-1\}$ is the set of $M$ or fewer indices $j$ for which the input regressor vectors $\textbf{u}(n-j)$ are orthogonal to each other. The orthogonalization process determines the indices forming the set $J_{n}$. The equation $\eqref{eq14}$ forms the basis for the performance analysis of the ZA-APA algorithm. Using the statistical independence between
$\textbf{w}_{2}(n)$ and $\textbf{u}(n)$ (i.e., ``independence assumption"),
and recalling that $\epsilon(n)$ is zero-mean i. i. d random variable which is independent of
$\textbf{u}(n)$ and thus of $\widetilde{\textbf{w}}_{2}(n)$, one can write
\begin{equation}\label{eq15}
\begin{split}
E[\widetilde{\textbf{w}}_{2}(n+1)]&=\bigg[\textbf{I}_{L}-\mu E\Big[  \sum\limits_{j\in J_{n}}^{} \frac{\textbf{u}(n-j) \textbf{u}^{T}(n-j)}{\textbf{u}^{T}(n-j)\textbf{u}(n-j)} \Big] \hspace{0.1em}\bigg] E[\widetilde{\textbf{w}}_{2}(n)]
+ \rho \hspace{0.2em} E\Big[ sgn\big(\textbf{w}_{2}(n)\big) \Big]
\end{split}
\end{equation}
Using A3), we can write the outer to inner product as,
\begin{equation}\label{eq16}
\begin{split}
\frac{\textbf{u}(n-j) \textbf{u}^{T}(n-j)}{\textbf{u}^{T}(n-j)\textbf{u}(n-j)}&= \frac{s(n-j)r(n-j)\mathbf{v}(n-j)\mathbf{v}^{T}(n-j)s(n-j)r(n-j)}{s^{2}(n-j)r^{2}(n-j) \|\mathbf{v}(n-j)\|^{2}}\\
& = \mathbf{v}(n-j)\mathbf{v}^{T}(n-j)\\
\end{split}
\end{equation}
where $\mathbf{v}(n-j) \in \{\textbf{\emph{v}}_{1},
\textbf{\emph{v}}_{2}, . . . , \textbf{\emph{v}}_{L}\}$. Note
that the above result is independent of the norm of $\textbf{u}(n-j)$. \\
Using the result presented in $\eqref{eq16}$, the Eq. $\eqref{eq15}$ can be rewritten as
\begin{equation}\label{eq17}
\begin{split}
E[\widetilde{\textbf{w}}_{2}(n+1)]&=\bigg[\textbf{I}_{L}-\mu E\Big[\sum\limits_{k\in K_{n}}^{}\textbf{\emph{v}}_{k} \textbf{\emph{v}}^{T}_{k}\Big] \hspace{0.1em}\bigg] E[\widetilde{\textbf{w}}_{2}(n)]
+ \rho \hspace{0.2em} E\Big[ sgn\big(\textbf{w}_{2}(n)\big) \Big]
\end{split}
\end{equation}
where
\begin{equation}\label{eq18}
\begin{split}
K_{n}&= \{k:\exists j\in J_{n}\ni\frac{\textbf{u}(n-j) \textbf{u}^{T}(n-j)}{\textbf{u}^{T}(n-j)\textbf{u}(n-j)}=\textbf{\emph{v}}_{k}\textbf{\emph{v}}^{T}_{k} \} \\
&\subseteq \{1, 2, \cdots, L\}
\end{split}
\end{equation}
By defining the vector $\boldsymbol{\alpha}(n)$ as the representation of $E[\widetilde{\textbf{w}}_{2}(n)]$ in terms of the orthonormal vectors $\{\textbf{\emph{v}}_{1}, \textbf{\emph{v}}_{2}, . . . , \textbf{\emph{v}}_{L}\}$. That is $\boldsymbol{\alpha}_{2}(n) \equiv \textbf{V}^{T} E[\widetilde{\textbf{w}}_{2}(n)] $ and $\boldsymbol{\alpha}_{2, i}(n)= \textbf{\emph{v}}^{T}_{i} E[\widetilde{\textbf{w}}_{2}(n)]$.\\
Using this notation, pre-multiplying $\eqref{eq17}$ by $\textbf{\emph{v}}^{T}_{i}$ results in
\begin{equation}\label{eq19}
\begin{split}
\textbf{\emph{v}}^{T}_{i} E[\widetilde{\textbf{w}}_{2}(n+1)]&=\bigg[\textbf{I}_{L}-\mu E\Big[\textbf{\emph{v}}^{T}_{i} \sum\limits_{k\in K_{n}}^{}\textbf{\emph{v}}_{k} \textbf{\emph{v}}^{T}_{k}\Big] \hspace{0.1em}\bigg] E[\widetilde{\textbf{w}}_{2}(n)]
+ \rho \hspace{0.2em} \textbf{\emph{v}}^{T}_{i} E\Big[ sgn\big(\textbf{w}_{2}(n)\big) \Big]
\end{split}
\end{equation}
From the orthonormality of $\textbf{\emph{v}}_{k}$'s we have
\begin{equation}\label{eq20}
\begin{split}
\textbf{\emph{v}}^{T}_{i} \sum\limits_{k\in K_{n}}^{}\textbf{\emph{v}}_{k}\textbf{\emph{v}}^{T}_{k}&= \begin{cases}
&\textbf{\emph{v}}^{T}_{i},\mspace{14mu} \text{if}\mspace{14mu} i \in K_{n}\\
&0,\mspace{27mu} \text{if}\mspace{14mu} i \notin K_{n}
\end{cases}
\end{split}
\end{equation}
Using the above result and substituting $P(i\in K_{n})=\beta_{i}$, Eq.\eqref{eq19} becomes
\begin{equation}\label{eq21}
\begin{split}
\alpha_{2, i}(n+1)&=[1-\mu \beta_{i}] \alpha_{2, i}(n)
+ \rho \hspace{0.2em} b_{i}(n)
\end{split}
\end{equation}
where $b_{i}(n)=\textbf{\emph{v}}^{T}_{i} E\Big[ sgn\big(\textbf{w}_{2}(n)\big) \Big]$ and the term $Pr(i\in K_{n})=\beta_{i}$ in $\eqref{eq21}$ is the
probability of drawing an eigen vector $\textbf{\emph{v}}_{i}$
from the eigen vectors set $\{\textbf{\emph{v}}_{1},
\textbf{\emph{v}}_{2}, . . . , \textbf{\emph{v}}_{L}\}$ at most in $M$ trials. If we assume $p_{i}$ is the probability of drawing
an eigen vector $\textbf{\emph{v}}_{i}$, then $Pr(i\in K_{n})= \beta_{i} = 1-(1-p_{i})^{M}$.\\
In steady state i.e., as $n \to \infty$, from \eqref{eq21}, it is possible to write
\begin{equation}\label{eq22}
\begin{split}
\lim\limits_{n \to \infty}\alpha_{2, i}(n)&= \frac{\rho}{\mu \beta_{i}} \hspace{0.2em} \lim\limits_{n \to \infty}b_{i}(n)
\end{split}
\end{equation}
To evaluate $\lim\limits_{n \to \infty}b_{i}(n)$,
we classify the unknown parameters (i.e., filter taps)
into two disjoint subsets same as in $\cite{13}$, denoting,
$NZ$ and $Z$ for active and inactive filter taps respectively.
If the $i^{th}$ tap of the optimum filter is active, then $i \in NZ$,
else $i \in Z$. For sufficiently small controller constant $\rho$,
for every $i \in NZ$, we have $|\widetilde{w}_{2, i}
(\infty)| \ll |w_{opt, i}|$. On the other hand for every $i \in Z$,
we have $|\widetilde{w}_{2, i}(\infty)| > |w_{opt, i}|$.
Since $w_{opt, i}=0$ for every $i \in Z$, so for zero mean Gaussian input,
in steady state we can assume $w_{2, i}(\infty)$ to be Gaussian with zero
mean.
\par
Therefore, from the aforementioned remarks and for the sufficiently small controller constant $\rho$, we can approximate,
\begin{equation}\label{eq23}
\begin{split}
\lim\limits_{n \to \infty} sgn(w_{2, i}(n))&=sgn(w_{opt, i}) \hspace{2em} \text{for} \hspace{0.5em} i \in NZ\\
\lim\limits_{n \to \infty} E\left[sgn(w_{2, i}(n))\right]&=0 \hspace{6.2em} \text{for} \hspace{0.5em} i \in Z
\end{split}
\end{equation}
These approximations are helpful to evaluate the expectations of non linear function $\lim\limits_{n \to \infty}b_{i}(n)$, involved in $\eqref{eq22}$ for both white and correlated inputs. To derive the closed form expressions for the steady state mean of the weight deviation ($\boldsymbol{\alpha}_{2, i}(\infty)$) of active and inactive coefficients, at this stage, we assume that the ZA-APA algorithm is operating on white input. This
assumption simplifies the approach and the eigen vectors set
$\{\textbf{\emph{v}}_{1}, \textbf{\emph{v}}_{2}, . . . , \textbf{\emph{v}}_{L}\}$
becomes the trivial basis $\{\textbf{\emph{e}}_{1},
\textbf{\emph{e}}_{2}, . . . , \textbf{\emph{e}}_{L}\}$.
Therefore, in steady state for the ZA-APA algorithm, we have,
\begin{equation}\label{eq24}
\begin{split}
\lim\limits_{n \to \infty} \alpha_{2, i}(n)= \lim\limits_{n \to \infty} E[\widetilde{w}_{2, i}(n)]=
\begin{cases}
 \frac{\rho}{\mu \beta_{i}}sgn(w_{opt, i}) \hspace{2em} \text{for} \hspace{0.5em} i \in NZ\\
0 \hspace{7.8em} \text{for} \hspace{0.5em} i \in Z
\end{cases}
\end{split}
\end{equation}
\subsection{Excess Mean Square Error Analysis of the ZA-APA Algorithm}
The EMSE of ZA-APA algorithm can be written in the following form,
\begin{equation}\label{eq25}
\begin{split}
J_{ex, 2}(n)&= E[e^{2}_{a, 2}(n)]=E[\widetilde{\textbf{w}}^{T}_{2}(n) \hspace{0.25em} \textbf{R} \hspace{0.25em}\widetilde{\textbf{w}}_{2}(n)]=Tr \left( \textbf{R} \hspace{0.25em} \textbf{K}_{2}(n) \right)= Tr \left( \boldsymbol{\Lambda} \hspace{0.25em} \textbf{V}^{T}\textbf{K}_{2}(n) \textbf{V}\right)
\end{split}
\end{equation}
where $E[\widetilde{\textbf{w}}_{2}(n) \widetilde{\textbf{w}}^{T}_{2}(n)]=\textbf{K}_{2}(n)$.
\par
Let us define the diagonal elements of the transformed covariance
matrix  $\textbf{V}^{T}\mspace{2mu}\textbf{K}_{2}(n)\mspace{2mu}\textbf{V}$ as $\widetilde{\lambda}_{2, i}(n)$
for $i=1, 2, . . . , L$. That is
\begin{equation}\label{eq26}
\begin{split}
[\textbf{V}^{T}\textbf{K}_{2}(n)\textbf{V}]_{ii}=\textbf{\emph{v}}_{i}^{T}\textbf{K}_{2}(n)\textbf{\emph{v}}_{i}=\widetilde{\lambda}_{2, i}(n)
\end{split}
\end{equation}
with the above notation EMSE of the ZA-APA algorithm can be written as follows:
\begin{equation}\label{eq27}
\begin{split}
J_{ex, 2}(n)&=\sum\limits_{i=1}^{L} \lambda_{i} \hspace{0.2em} \widetilde{\lambda}_{2, i}(n)
\end{split}
\end{equation}
Using the statistical independence between
$\textbf{w}_{1}(n)$ and $\textbf{u}(n)$ (i.e., ``independence assumption"),
and recalling that $\varepsilon(n)$ is of zero-mean and also independent of
$\textbf{u}(n)$ and thus of $\widetilde{\textbf{w}}_{1}(n)$, one can write the recursion for the mean square deviation of ZA-APA algorithm as follows:
\begin{equation*}\label{eq28}
\begin{split}
\textbf{K}_{2}(n+1)  =
\textbf{K}_{2}(n) &- \mspace{4mu} \mu \mspace{4mu} E \left[\sum\limits_{j \in J_{n}}^{}\frac{\textbf{u}(n-j) \textbf{u}^{T}(n-j)}{\textbf{u}^{T}(n-j)\textbf{u}(n-j)}\right] \mspace{4mu} \textbf{K}_{2}(n)- \mspace{4mu} \mu \mspace{4mu} \textbf{K}_{2}(n) \mspace{4mu} E\left[\sum\limits_{m\in J_{n}}^{}\frac{\textbf{u}(n-m) \textbf{u}^{T}(n-m)}{\textbf{u}^{T}(n-m)\textbf{u}(n-m)}\right]\\
& +  \mu^{2} \mspace{4mu} E\left[\left(\sum\limits_{j \in J_{n}}^{}\frac{\textbf{u}(n-j) \textbf{u}^{T}(n-j)}{\textbf{u}^{T}(n-j)\textbf{u}(n-j)}\right) \mspace{4mu} \textbf{K}_{2}(n) \left(\sum\limits_{m \in J_{n}}^{}\frac{\textbf{u}(n-m) \textbf{u}^{T}(n-m)}{\textbf{u}^{T}(n-m)\textbf{u}(n-m)}\right)\right]\\
& +  \mu^{2} \mspace{4mu} E\left[\left(\sum\limits_{j \in J_{n}}^{}\frac{\textbf{u}(n-j) \varepsilon(n-j)}{\textbf{u}^{T}(n-j)\textbf{u}(n-j)}\right) \mspace{4mu}  \left(\sum\limits_{m \in J_{n}}^{}\frac{\textbf{u}(n-m) \varepsilon(n-m)}{\textbf{u}^{T}(n-m)\textbf{u}(n-m)}\right)^{T} \right]\\
\end{split}
\end{equation*}
\begin{align}\label{eq28}
\begin{split}
& +  \rho \mspace{4mu} E\left[I_{L} - \mu \mspace{4mu} \left(\sum\limits_{j \in J_{n}}^{}\frac{\textbf{u}(n-j) \textbf{u}^{T}(n-j)}{\textbf{u}^{T}(n-j)\textbf{u}(n-j)}\right)\right]\mspace{4mu} E\left[\widetilde{\textbf{w}}_{2}(n)\mspace{4mu}sgn(\textbf{w}^{T}_{2}(n))\right]\\
& + \rho \mspace{4mu} E\left[sgn(\textbf{w}_{2}(n))\mspace{4mu} \widetilde{\textbf{w}}^{T}_{2}(n)\right] \mspace{2mu} E\left[I_{L} - \mu \mspace{4mu} \left(\sum\limits_{m \in J_{n}}^{}\frac{\textbf{u}(n-m) \textbf{u}^{T}(n-m)}{\textbf{u}^{T}(n-m)\textbf{u}(n-m)}\right)\right]\\
& +  \rho^{2} \mspace{4mu}  E\left[sgn\left(\textbf{w}_{2}(n)\right)\mspace{4mu} sgn\left(\textbf{w}^{T}_{2}(n)\right)\right]
\end{split}
\end{align}
Using $\eqref{eq18}$ to replace the outer-to-inner product ratios with $\textbf{\emph{v}}_{k}\textbf{\emph{v}}^{T}_{k}$, one can have
\begin{equation}\label{eq29}
\begin{split}
\textbf{K}_{2}(n+1)&= \textbf{K}_{2}(n)-\mspace{2mu}\mu \mspace{2mu} E\left[\sum\limits_{k\in K_{n}}^{}\textbf{\emph{v}}_{k} \textbf{\emph{v}}^{T}_{k}\right]\mspace{2mu}\textbf{K}_{2}(n)\mspace{2mu}-\mu \mspace{2mu}\textbf{K}_{2}(n)\mspace{2mu}E\left[\sum\limits_{m\in K_{n}}^{}\textbf{\emph{v}}_{m} \textbf{\emph{v}}^{T}_{m}\right]\\
&\hspace{1em}+\mu^{2}\mspace{2mu}E\left[\left(\sum\limits_{k\in K_{n}}^{}\textbf{\emph{v}}_{k}\textbf{\emph{v}}^{T}_{k}\right)\textbf{K}_{2}(n)
\left(\sum\limits_{m\in K_{n}}^{}\textbf{\emph{v}}_{m} \textbf{\emph{v}}^{T}_{m}\right)\right]+\mspace{2mu}\mu^{2}\xi^{0}E[\frac{1}{r^{2}}]\mspace{2mu}E\left[\sum\limits_{k\in K_{n}}^{}\textbf{\emph{v}}_{k}\textbf{\emph{v}}^{T}_{k}\right]\\
&\hspace{1em}+\mspace{2mu}\rho\mspace{2mu}E\left[I_{L}-\mu\mspace{2mu}\sum\limits_{k\in K_{n}}^{}\textbf{\emph{v}}_{k}\textbf{\emph{v}}^{T}_{k}\right]\mspace{4mu}E\left[\widetilde{\textbf{w}}_{2}(n)\mspace{4mu}sgn(\textbf{w}^{T}_{2}(n))\right]\\
&\hspace{1em}+\mspace{2mu}\rho\mspace{2mu} E\left[sgn(\textbf{w}_{2}(n))\mspace{2mu}\widetilde{\textbf{w}}^{T}_{2}(n)\right] \mspace{4mu} E\left[I_{L}-\mu\mspace{2mu}\sum\limits_{m\in K_{n}}^{}\textbf{\emph{v}}_{m}\textbf{\emph{v}}^{T}_{m}\right]\\
&\hspace{1em} + \mspace{4mu} \rho^{2} \mspace{4mu}  E\left[sgn\left(\textbf{w}_{2}(n)\right)\mspace{4mu} sgn\left(\textbf{w}^{T}_{2}(n)\right)\right]
\end{split}
\end{equation}
With the notation presented in $\eqref{eq26}$, pre and post multiplication
of \eqref{eq29} by $\textbf{\emph{v}}_{i}^{T}$ and
$\textbf{\emph{v}}_{i}$ respectively, results in
\begin{equation}\label{eq30}
\begin{split}
\widetilde{\lambda}_{2, i}(n+1)&=\widetilde{\lambda}_{2, i}(n)-\mu\mspace{2mu}E\left[\textbf{\emph{v}}_{i}^{T}\left(\sum\limits_{k\in K_{n}}^{}\textbf{\emph{v}}_{k}\textbf{\emph{v}}^{T}_{k}\right) \right]\mspace{2mu}\textbf{K}_{2}(n)\mspace{2mu}\textbf{\emph{v}}_{i}
-\mu\mspace{2mu}\textbf{\emph{v}}_{i}^{T}\mspace{2mu}\textbf{K}_{2}(n)\mspace{2mu}E\left[\left(\sum\limits_{m\in K_{n}}^{}\textbf{\emph{v}}_{m}\textbf{\emph{v}}^{T}_{m}\right) \textbf{\emph{v}}_{i}\right]\\
&\hspace{1em}+\mu^{2}E\left[\textbf{\emph{v}}_{i}^{T}\mspace{-8mu}\left(\sum\limits_{k\in K_{n}}^{}\mspace{-4mu}\textbf{\emph{v}}_{k}\textbf{\emph{v}}^{T}_{k}\mspace{-8mu}\right) \textbf{K}_{2}(n)\mspace{-8mu}\left(\sum\limits_{m\in K_{n}}^{}\textbf{\emph{v}}_{m}\textbf{\emph{v}}^{T}_{m}\right)\mspace{-5mu}\textbf{\emph{v}}_{i}\right]
+\mu^{2}\mspace{2mu}\xi^{0}E[\frac{1}{r^{2}}]E\left[\textbf{\emph{v}}_{i}^{T}\left( \sum\limits_{k\in K_{n}}^{}\textbf{\emph{v}}_{k}\textbf{\emph{v}}^{T}_{k}\right)\textbf{\emph{v}}_{i}\right]\\
&\hspace{1em}+\rho\mspace{2mu}E\left[\textbf{\emph{v}}_{i}^{T} \left( I_{L}-\mu\sum\limits_{k\in K_{n}}^{}\textbf{\emph{v}}_{k}\textbf{\emph{v}}^{T}_{k} \right)\right]\mspace{4mu} E\left[\left(\widetilde{\textbf{w}}_{2}(n)\mspace{2mu}sgn(\textbf{w}^{T}_{2}(n))\right)   \textbf{\emph{v}}_{i}\right]\\
&\hspace{1em}+\rho\mspace{2mu}E\left[\textbf{\emph{v}}_{i}^{T} \left(sgn(\textbf{w}_{2}(n))\mspace{2mu} \widetilde{\textbf{w}}^{T}_{2}(n)\right)              \right]\mspace{4mu} E\left[\left( I_{L}-\mu\sum\limits_{m\in K_{n}}^{}\textbf{\emph{v}}_{m}\textbf{\emph{v}}^{T}_{m} \right)\textbf{\emph{v}}_{i}\right]\\
& \hspace{1em}+ \mspace{4mu} \rho^{2} \mspace{4mu}  E\left[\textbf{\emph{v}}_{i}^{T}\left(sgn\left(\textbf{w}_{2}(n)\right)\mspace{4mu} sgn\left(\textbf{w}^{T}_{2}(n)\right)\right)\textbf{\emph{v}}_{i}\right]
\end{split}
\end{equation}
From the orthonormality of $\textbf{\emph{v}}_{k}$'s,
using the result presented in $\eqref{eq20}$ and substituting $P(i\in K_{n})=\beta_{i}$, Eq.\eqref{eq30} becomes
\begin{equation}\label{eq31}
\begin{split}
\widetilde{\lambda}_{2, i}(n+1)&= \underbrace{ [1-\mu(2-\mu)\beta_{i}] \hspace{0.5em}\widetilde{\lambda}_{2, i}(n) + \mu^{2} \xi^{0} E[\frac{1}{r^{2}}]\beta_{i} }_\text{$\widetilde{\lambda}_{1, i}(n+1)$ of APA $\cite{18}$}  \\
&+ \underbrace{ \rho\mspace{2mu}(1-\mu\beta_{i})\hspace{0.5em}\left[\textbf{\emph{v}}^{T}_{i} \hspace{0.5em} \boldsymbol{\Phi}(n) \hspace{0.5em}\textbf{\emph{v}}_{i}\right] + \rho\mspace{2mu}(1-\mu\beta_{i})\hspace{0.5em}\left[\textbf{\emph{v}}^{T}_{i} \hspace{0.5em} \boldsymbol{\Phi}^{T}(n) \hspace{0.5em}\textbf{\emph{v}}_{i}\right] + \mspace{2mu}\rho^{2}\mspace{2mu} \left[\textbf{\emph{v}}^{T}_{i} \hspace{0.5em} \boldsymbol{\Psi}(n) \hspace{0.5em}\textbf{\emph{v}}_{i}\right] }_\text{$\Delta_{2, i}(n)$}
\end{split}
\end{equation}
where
$\boldsymbol{\Phi}(n)= E\left[ \widetilde{\textbf{w}}_{2}(n)\mspace{2mu}sgn(\textbf{w}^{T}_{2}(n)) \right]$  and
$\boldsymbol{\Psi}=E\left[sgn(\textbf{w}_{2}(n))\mspace{4mu} sgn(\textbf{w}^{T}_{2}(n))\right]$
\par
In steady state i.e., as $n \to \infty$, from \eqref{eq31}, it is possible to write
\begin{equation}\label{eq32}
\begin{split}
\widetilde{\lambda}_{2, i}(\infty)&= \underbrace{ \frac{\mu}{2-\mu}  \xi^{0} E[\frac{1}{r^{2}}] }_\text{$\widetilde{\lambda}_{1, i}(\infty)$ of APA $\cite{18}$} \\
&+ \underbrace{ \frac{\rho\mspace{2mu}(1-\mu\beta_{i})}{\mu (2-\mu)\beta_{i}}\hspace{0.5em}\left[\textbf{\emph{v}}^{H}_{i} \hspace{0.5em} \boldsymbol{\Phi}(\infty) \hspace{0.5em}\textbf{\emph{v}}_{i}\right] + \frac{\rho\mspace{2mu}(1-\mu\beta_{i})}{\mu (2-\mu)\beta_{i}}\hspace{0.5em}\left[\textbf{\emph{v}}^{T}_{i} \hspace{0.5em} \boldsymbol{\Phi}^{T}(\infty) \hspace{0.5em}\textbf{\emph{v}}_{i}\right] + \mspace{2mu}\frac{\rho^{2}}{\mu (2-\mu)\beta_{i}}\mspace{2mu} \left[\textbf{\emph{v}}^{T}_{i} \hspace{0.5em} \boldsymbol{\Psi}(\infty) \hspace{0.5em}\textbf{\emph{v}}_{i}\right] }_\text{$\Delta_{2, i}(\infty)$}
\end{split}
\end{equation}
where
$\boldsymbol{\Phi}(\infty)= \lim\limits_{n \to \infty} E\left[ \widetilde{\textbf{w}}_{2}(n)\mspace{2mu}sgn(\textbf{w}^{T}_{2}(n))  \right]$  and
$\boldsymbol{\Psi}(\infty)=\lim\limits_{n \to \infty} E\left[sgn(\textbf{w}_{2}(n))\mspace{4mu} sgn(\textbf{w}^{T}_{2}(n))\right]$
\par
To evaluate $\boldsymbol{\Phi}(\infty)$ and $\boldsymbol{\Psi}(\infty)$, as mentioned in the convergence in mean section
the unknown parameters (i.e., filter taps) are classified
into two disjoint subsets denoting,
$NZ$ and $Z$ for active and inactive filter taps respectively.
Under small misadjustment condition ( i.e., the steady state standard deviation of $w_{2, i}$ ( $\sigma_{w_{2, i}}$) is very small), also following the lines of $\cite{22}$, we can approximate $\lim\limits_{n \to \infty} E[\widetilde{w}_{2, i}(n) sgn(w_{2, j}(n))] = \lim\limits_{n \to \infty} E[\widetilde{w}_{2, i}(n)] E[sgn(w_{2, j}(n))]$  and $\lim\limits_{n \to \infty} E[sgn(w_{2, i}(n))$ $ sgn(w_{2, j}(n))] =  \lim\limits_{n \to \infty} E[sgn(w_{2, i}(n))] E[sgn(w_{2, j}(n))]$ for all $i \neq j$. These approximations are helpful to evaluate the expectations of non linear functions involved in $\eqref{eq32}$ for both white and correlated input.
To derive the closed form expressions for the steady state mean
square deviation ($\widetilde{\lambda}_{2, i}(\infty)$) of
active and inactive coefficients, we assume
that the ZA-APA algorithm is operating on white input. This
assumption simplifies the approach and the eigen vectors set
$\{\textbf{\emph{v}}_{1}, \textbf{\emph{v}}_{2}, . . . , \textbf{\emph{v}}_{L}\}$
becomes the trivial basis $\{\textbf{\emph{e}}_{1},
\textbf{\emph{e}}_{2}, . . . , \textbf{\emph{e}}_{L}\}$.
Therefore, in steady state for the ZA-APA algorithm, we have,
\begin{equation}\label{eq33}
\begin{split}
\boldsymbol{\Phi}_{i, i}(\infty)=
\begin{cases}
E\left[ \widetilde{w}_{2, i}(\infty) \right] \mspace{2mu}sgn(w_{opt, i}) \mspace{129mu} \text{for }\mspace{5mu} i \in NZ\\
E\left[ \widetilde{w}_{2, i}(\infty)\mspace{2mu}sgn(w_{2, i}(\infty)) \right]    \mspace{110mu} \text{for }\mspace{5mu} i \in Z
\end{cases}
\end{split}
\end{equation}
and
\begin{equation}\label{eq34}
\begin{split}
\boldsymbol{\Psi}_{i, i}(\infty)=
\begin{cases}
E[\mspace{4mu} \left[ sgn(w_{opt, i} ) \right]^{2} \mspace{4mu}]&=1    \mspace{185mu} \text{for }\mspace{5mu} i \in NZ\\
E[\mspace{4mu} \left[ sgn(w_{2, i}(\infty)) \right]^{2} \mspace{4mu}]&=1      \mspace{190mu} \text{for }\mspace{5mu} i \in Z\\
\end{cases}
\end{split}
\end{equation}
Using these results for every $i \in NZ$ we can write,
\begin{equation}\label{eq35}
\begin{split}
\widetilde{\lambda}_{2, i}(\infty)&= \underbrace{ \frac{\mu}{2-\mu}  \xi^{0} E[\frac{1}{r^{2}}] }_\text{$\widetilde{\lambda}_{1, i}(\infty)$ } + \hspace{1em}\underbrace{ \frac{\rho^{2} (2-\mu \beta_{i})}{\mu^{2}\beta^{2}_{i}[2-\mu]} }_\text{$\Delta_{2, i}(\infty)$}
\end{split}
\end{equation}
On the other hand for $i \in Z$, $\widetilde{w}_{2, i}(\infty)=
w_{opt, i} - w_{2, i}(\infty)= - w_{2, i}(\infty)$. Assuming $w_{1, i}(\infty)$and
$w_{2, i}(\infty)$ to be jointly Gaussian (having mean zero in the steady state,
as $w_{opt, i}(\infty)=0$), using Price's theorem we can write, $E\left[sgn(w_{2, i}(\infty))\mspace{2mu}
\widetilde{w}_{2, i}(\infty)\right]=-E\left[sgn(w_{2, i}(\infty))
\mspace{2mu}{w}_{2, i}(\infty)\right]=-gE\left[w_{2, i}(\infty)
\mspace{2mu}w_{2, i}(\infty)\right]=-gE\left[\widetilde{w}_{2, i}
(\infty)\mspace{2mu}\widetilde{w}_{2, i}(\infty)\right]$,
where $g=\sqrt{\frac{2}{\pi\mspace{1mu}\sigma^{2}_{w_{2, i}}}}=\sqrt{\frac{2}{\pi\mspace{1mu}\widetilde{\lambda}_{2, i}(\infty)}}$.
\par
With these results for $i \in Z$, the equation $\eqref{eq32}$ can be written as,
\begin{equation}\label{eq36}
\begin{split}
\widetilde{\lambda}_{2, i}(\infty)&= \frac{ \mu^{2} \beta_{i} \xi^{0} E[\frac{1}{r^{2}}]\mspace{4mu} + \rho^{2} }{ \mu (2-\mu) \beta_{i} + 2 \rho (1-\mu \beta_{i}) \sqrt{\frac{2}{\pi} }  \sqrt{\frac{1}{\small{\widetilde{\lambda}_{2, i}(\infty)}}} }
\end{split}
\end{equation}
the above equation is a quadratic of the form,
\begin{subequations}\label{eq37}
    \begin{equation}
  a \mspace{4mu} \widetilde{\lambda}_{1, i}(\infty) + b \mspace{4mu} \sqrt{\widetilde{\lambda}_{1, i}(\infty)} + c =0
   \end{equation}
   \text{ where the coefficients are,}
   \begin{equation}
    \begin{split}
  & a= 1\\
  & b= \sqrt{\frac{8}{\pi}}\mspace{4mu} \frac{\rho (1-\mu \beta_{i})}{\mu (2-\mu)\beta_{i}} \\
  & c= -\left( \frac{\mu}{2-\mu}\xi^{0} E[\frac{1}{r^{2}}] +  \frac{\rho^{2}}{\mu (2-\mu)\beta_{i}} \right)
  \end{split}
  \end{equation}
    \end{subequations}
Then, for $i \in Z$, we have ( since $\sigma_{w_{2, i}}$ is
positive, note that only real, positive root has to be considered.)
\begin{equation}\label{eq38}
\begin{split}
\sqrt{\widetilde{\lambda}_{2, i}(\infty)}& = \frac{1}{\mu (2-\mu) \beta_{i}} \left[-\sqrt{\frac{2}{\pi}} \rho (1-\mu \beta_{i})+ \sqrt{\frac{2}{\pi} \rho^{2} (1-\mu \beta_{i})^2 + \left(\mu^{2} \beta_{i} \xi^{0} E[\frac{1}{r^{2}}] + \rho^{2}\right) \mu (2-\mu) \beta_{i} } \mspace{4mu} \right]
\end{split}
\end{equation}
Squaring both the sides of the above equation, the steady state mean
square deviation of single inactive tap can be written as,
\begin{equation}\label{eq39}
\begin{split}
\widetilde{\lambda}_{2, i}(\infty)&= \underbrace{ \widetilde{\lambda}_{1, i}(\infty)  }_\text{ APA}
 \hspace{1em} + \hspace{1em} \underbrace{  \left[ \begin{array}{l} \hspace{1em} \rho^{2} \left( \frac{\frac{4}{\pi} (1- \mu \beta_{i})^{2} + \mu (2-\mu) \beta_{i} } { \mu^{2} (2-\mu)^{2} \beta^{2}_{i} } \right) \\ \hspace{0.5em} - \sqrt{\frac{8}{\pi}} \rho \frac {(1-\mu \beta_{i}) }{\mu (2-\mu) \beta_{i} } \sqrt{ \rho^{2} \left( \frac{\frac{2}{\pi} (1-\mu \beta_{i})^{2} + (\mu (2-\mu)\beta_{i}) }{\mu^{2}(2-\mu)^{2}\beta^{2}_{i}}      \right)  + \widetilde{\lambda}_{1, i}(\infty)}  \end{array} \right] }_\text{$\Delta_{1, i}(\infty)$}\\
 &= \widetilde{\lambda}_{1, i}(\infty) - \sqrt{\frac{8}{\pi}} \hspace{0.5em}  \frac {\rho(1-\mu \beta_{i}) }{\mu (2-\mu) \beta_{i} } \sqrt{ \widetilde{\lambda}_{1, i}(\infty) }  \hspace{5em}  \rho^{2} \ll \rho
\end{split}
\end{equation}
From the above equation, it can be observed that for having
$\widetilde{\lambda}_{2, i}(\infty) \leq \widetilde{\lambda}_{1, i}(\infty)$,
the controller constant $\rho$ has to be chosen very carefully.
By using $\eqref{eq35}$ and $\eqref{eq39}$ EMSE of ZA-APA can be calculated.
\subsubsection{$\rho$ Range}
For a system with given length $L$ and projection order $M$, $J_{ex, 2}(\infty) \leq J_{ex, 1}(\infty)$, if and only if
\begin{equation}\label{eq40}
\begin{split}
\rho^{2} \leq \frac{8}{\pi}\frac{(L-K)^{2} \hspace{0.25em} (1-\mu\beta{i})^{2} \hspace{0.25em}  \mu^{3}\beta^{2}_{i}\xi^{0}E[\frac{1}{r^{2}}] }{C}
\end{split}
\end{equation}
where \\
$C= \left( \hspace{0.45em} K(L-K) \left( 2-\mu\beta_{i}\right) \left( \frac{8}{\pi} (1-\mu\beta_{i})^{2} + 2 \mu (2-\mu)\beta_{i} \right) \hspace{0.45em} \right)
+ \left(\hspace{0.45em} K^{2}(2-\mu\beta_{i})^{2} (2-\mu) \hspace{0.45em}\right) \\
-\left ( \hspace{0.45em}  (L-k)^{2} (2-\mu) \mu^{2}\beta_{i}^{2}\hspace{0.45em}\right)$
\par
This bound shows the dependence of regularization parameter $\rho$ value on the projection order $M$.
\subsection{Cross Excess Mean Square Error Analysis of ZA-APA and APA Algorithms}
The cross EMSE of the proposed combination can be written in the following form,
\begin{equation}\label{eq41}
\begin{split}
J_{ex, 12}(n)&= E[\widetilde{\textbf{w}}^{T}_{2}(n) \hspace{0.25em} \textbf{R} \hspace{0.25em}\widetilde{\textbf{w}}_{1}(n)]=Tr \left( \textbf{R} \hspace{0.25em} \textbf{K}_{12}(n) \right)= Tr \left( \boldsymbol{\Lambda} \hspace{0.25em} \textbf{V}^{T} \hspace{0.25em}\textbf{K}_{12}(n) \hspace{0.25em} \textbf{V} \right)
\end{split}
\end{equation}
where $E[\widetilde{\textbf{w}}_{1}(n) \widetilde{\textbf{w}}^{T}_{2}(n)]=\textbf{K}_{12}(n)$.
\par
By defining the diagonal elements of the transformed covariance
matrix  $\textbf{V}^{T}K_{12}(n)\textbf{V}$ as $\widetilde{\lambda}_{12, i}(n)$ for $i=1, 2, . . . , L$. That is
\begin{equation}\label{eq42}
\begin{split}
[\textbf{V}^{T}K_{12}(n)\textbf{V}]_{ii}=\textbf{\emph{v}}_{i}^{T}K_{12}(n)\textbf{\emph{v}}_{i}=\widetilde{\lambda}_{12, i}(n)
\end{split}
\end{equation}
With the above notation cross EMSE of the combination can be written as follows:
\begin{equation}\label{eq43}
\begin{split}
J_{ex, 12}(n)&= \sum\limits_{i=1}^{L} \lambda_{i} \hspace{0.25em} \widetilde{\lambda}_{12, i}(n)
\end{split}
\end{equation}
Post multiplying $\widetilde{\textbf{w}}_{1}(n+1)$ by $\widetilde{\textbf{w}}^{T}_{2}(n+1)$,
taking expectation, and using assumptions A2) and A3) we get
\begin{equation}\label{eq44}
\begin{split}
\textbf{K}_{12}(n+1)&= \textbf{K}_{12}(n)-\mspace{2mu}\mu \mspace{2mu} E\left[\sum\limits_{k\in K_{n}}^{}\textbf{\emph{v}}_{k} \textbf{\emph{v}}^{T}_{k}\right]\mspace{2mu}\textbf{K}_{12}(n)-\mspace{2mu}\mu \mspace{2mu}\textbf{K}_{12}(n)\mspace{2mu}E\left[\sum\limits_{m\in K_{n}}^{}\textbf{\emph{v}}_{m} \textbf{\emph{v}}^{T}_{m}\right]\\
&+\mspace{2mu}\mu^{2}\mspace{2mu}E[\left(\sum\limits_{k\in K_{n}}^{}\textbf{\emph{v}}_{k}\textbf{\emph{v}}^{T}_{k}\right)\textbf{K}_{12}(n)
\left(\sum\limits_{m \in K_{n}}^{}\textbf{\emph{v}}_{m} \textbf{\emph{v}}^{T}_{m}\right)]+\mspace{2mu}\mu^{2}\xi^{0}E[\frac{1}{r^{2}}]\mspace{2mu}E\left[\sum\limits_{k\in K_{n}}^{}\textbf{\emph{v}}_{k}\textbf{\emph{v}}^{T}_{k}\right]\\
&+\mspace{2mu}\rho\mspace{4mu}   E\left[I_{L}-\mu\mspace{2mu}\sum\limits_{k\in K_{n}}^{}\textbf{\emph{v}}_{k}\textbf{\emph{v}}^{T}_{k}\right]  \mspace{4mu} E\left[\widetilde{\textbf{w}}_{1}(n) \hspace{0.1em} sgn(\textbf{w}^{T}_{2}(n))\right]
\end{split}
\end{equation}
With the above notation, the pre and post multiplication
of \eqref{eq44} by $\textbf{\emph{v}}_{i}^{T}$ and $\textbf{\emph{v}}_{i}$ respectively, results in
\begin{equation}\label{eq45}
\begin{split}
\widetilde{\lambda}_{12, i}(n+1)&=\widetilde{\lambda}_{12, i}(n)-\mu\mspace{2mu}E\left[\textbf{\emph{v}}_{i}^{T}\left(\sum\limits_{k\in K_{n}}^{}\textbf{\emph{v}}_{k}\textbf{\emph{v}}^{T}_{k}\right) \right]\mspace{2mu}\textbf{K}_{12}(n)\mspace{2mu}\textbf{\emph{v}}_{i} -\mu\mspace{2mu}\textbf{\emph{v}}_{i}^{T}\mspace{2mu}\textbf{K}_{12}(n)\mspace{2mu}E\left[\left(\sum\limits_{k\in K_{n}}^{}\textbf{\emph{v}}_{k}\textbf{\emph{v}}^{T}_{k}\right) \textbf{\emph{v}}_{i}\right]\\
&+\mu^{2}E\left[\textbf{\emph{v}}_{i}^{T}\mspace{-8mu}\left(\sum\limits_{k\in K_{n}}^{}\mspace{-4mu}\textbf{\emph{v}}_{k}\textbf{\emph{v}}^{T}_{k}\mspace{-8mu}\right) \textbf{K}_{12}(n)\mspace{-8mu}\left(\sum\limits_{m\in K_{n}}^{}\textbf{\emph{v}}_{m}\textbf{\emph{v}}^{T}_{m}\right)\mspace{-5mu}\textbf{\emph{v}}_{i}\right]
+\mu^{2}\mspace{2mu}\xi^{0}E[\frac{1}{r^{2}}]E\left[\textbf{\emph{v}}_{i}^{T}\left( \sum\limits_{k\in K_{n}}^{}\textbf{\emph{v}}_{k}\textbf{\emph{v}}^{T}_{k}\right)\textbf{\emph{v}}_{i}\right]\\
&+\rho\mspace{2mu}   E\left[\textbf{\emph{v}}_{i}^{T} \left( I_{L}-\mu\sum\limits_{k\in K_{n}}^{}\textbf{\emph{v}}_{k}\textbf{\emph{v}}^{T}_{k} \right)\right]  E\left[\widetilde{\textbf{w}}_{1}(n) \hspace{0.1em} sgn(\textbf{w}^{T}_{2}(n)) \right]   \textbf{\emph{v}}_{i}
\end{split}
\end{equation}
From the orthonormality of $\textbf{\emph{v}}_{k}$'s and
substituting $Pr(i\in K_{n})=\beta_{i}$, Eq. $\eqref{eq45}$ can be written as,
\begin{equation}\label{eq46}
\begin{split}
\widetilde{\lambda}_{12, i}(n+1)&= \underbrace{ \widetilde{\lambda}_{12, i}(n)[1-\mu(2-\mu)\beta_{i}] + \mu^{2} \xi^{0} E[\frac{1}{r^{2}}]\beta_{i} }_\text{$\widetilde{\lambda}_{1, i}(n+1)$} +\underbrace{ \rho[1-\mu\beta_{i}]E\left[\textbf{\emph{v}}^{T}_{i} \Big(\widetilde{\textbf{w}}_{1}(n) \hspace{0.1em} sgn(\textbf{w}^{T}_{2}(n)) \Big) \textbf{\emph{v}}_{i}\right] }_\text{$ \Delta_{12, i}(n)$}
\end{split}
\end{equation}
In steady state i.e., as $n \to \infty$, the term
$\widetilde{\lambda}_{12, i}(n)$ becomes time invariant.
Therefore, we have
\begin{equation}\label{eq47}
\begin{split}
\widetilde{\lambda}_{12, i}(\infty)&= \underbrace{ \frac{\mu}{2-\mu}\mspace{4mu}\xi^{0}\mspace{4mu}E[\frac{1}{r^{2}}] }_\text{$\widetilde{\lambda}_{1, i}(\infty)$} + \underbrace{  \frac{\rho(1-\mu\beta_{i})}{\mu(2-\mu)\beta_{i}} E\left[\textbf{\emph{v}}^{T}_{i} \Big(\widetilde{\textbf{w}}_{1}(\infty) \hspace{0.1em} sgn(\textbf{w}^{T}_{2}(\infty)) \Big) \textbf{\emph{v}}_{i}\right]  }_\text{$\Delta_{12, i}(\infty) $}
\end{split}
\end{equation}
Using the same approximations used in EMSE analysis of ZA-APA,
we can solve the term $ E\left[\textbf{\emph{v}}^{T}_{i} \Big(\widetilde{\textbf{w}}_{1}(\infty) \hspace{0.1em} sgn(\textbf{w}^{T}_{2}(\infty)) \Big) \textbf{\emph{v}}_{i}\right] $ in the above equation for
white and correlated input. However, to derive the closed
form expressions for both active and inactive coefficients,
we are assuming that the algorithms are operating on white input.
\par
Using the fact that $\lim\limits_{n \to \infty} E\left[\widetilde{w}_{1, i}(n)\right]=0$, for every $i \in NZ$ one can write,
\begin{equation}\label{eq48}
\begin{split}
\Delta_{12, i}(\infty) &=  \frac{\rho(1-\mu\beta_{i})}{\mu(2-\mu)\beta_{i}} E\left[\widetilde{\textbf{w}}_{1, i}(\infty) \hspace{0.1em} sgn(\textbf{w}_{2, i}(\infty)) \right]  \\
&= \frac{\rho(1-\mu\beta_{i})}{\mu(2-\mu)\beta_{i}}  E\left[ \widetilde{w}_{1, i}(\infty)\mspace{2mu} sgn(w_{opt, i}(\infty)) \right]  = 0
\end{split}
\end{equation}
and using Price's theorem for every $i \in Z$ we can write,
\begin{equation}\label{eq49}
\begin{split}
\Delta_{12, i}(\infty) &=  \frac{\rho(1-\mu\beta_{i})}{\mu(2-\mu)\beta_{i}} E\left[\widetilde{\textbf{w}}_{1, i}(\infty) \hspace{0.1em} sgn(\textbf{w}_{2, i}(\infty)) \right]  \\
&= - \frac{\rho(1-\mu\beta_{i})}{\mu(2-\mu)\beta_{i}} \mspace{2mu} \sqrt{ \frac{2}{\pi \sigma^{2}_{w_{2, i}} } } \mspace{10mu} \widetilde{\lambda}_{12, i}(\infty)\\
&= - \frac{\rho(1-\mu\beta_{i})}{\mu(2-\mu)\beta_{i}} \mspace{2mu} \sqrt{ \frac{2}{\pi } } \mspace{10mu} \frac{\widetilde{\lambda}_{12, i}(\infty)}{ \sqrt{\widetilde{\lambda}_{2, i}(\infty)} }
\end{split}
\end{equation}
From $\eqref{eq47}$, for every $i \in NZ$ the mean square deviation can be written as,
\begin{equation}\label{eq50}
\begin{split}
\widetilde{\lambda}_{12, i}(\infty)&= \underbrace{ \frac{\mu}{2-\mu}\mspace{4mu}\xi^{0}\mspace{4mu}E[\frac{1}{r^{2}}] }_\text{$\widetilde{\lambda}_{1, i}(\infty)$}
\end{split}
\end{equation}
and for $i \in Z$,
\begin{equation}\label{eq51}
\begin{split}
[\mu(2-\mu)\beta_{i}]\mspace{2mu}\widetilde{\lambda}_{12, i}(\infty)&= \mu^{2} \xi^{0} \beta_{i}  E[\frac{1}{r^{2}}]
-\rho (1-\mu\beta_{i}) \sqrt{\frac{2}{\pi }}\mspace{1mu}\frac{\widetilde{\lambda}_{12, i}(\infty)}{ \sqrt{\widetilde{\lambda}_{2, i}(\infty)} }
\end{split}
\end{equation}
By rearranging the terms in the above equation, we get
\begin{equation}\label{eq52}
\begin{split}
\widetilde{\lambda}_{12, i}(\infty)= \frac{\mu^{2} \xi^{0} \beta_{i} E[\frac{1}{r^{2}}]}{\left(\left[\mu(2-\mu)\beta_{i}\right] + \left[\rho (1-\mu\beta_{i})\sqrt{\frac{2}{\pi }} \frac{1}{ \sqrt{\widetilde{\lambda}_{2, i}(\infty)} }\right] \right)}
\end{split}
\end{equation}
From  $\eqref{eq38}$ and $\eqref{eq50}$
\begin{equation}\label{eq53}
\begin{split}
\widetilde{\lambda}_{12, i}(\infty)&= \hspace{1em}\frac{\mu^{2} \xi^{0} \beta_{i} E[\frac{1}{r^{2}}]  \sqrt{\widetilde{\lambda}_{2, i}(\infty)} }{\left(\left[\mu(2-\mu)\beta_{i} \sqrt{ \widetilde{\lambda}_{2, i}(\infty) }  \right] + \left[\rho (1-\mu\beta_{i})\mspace{1mu} \sqrt{\frac{2}{\pi }}\mspace{5mu} \right ] \right)}\\
&= \hspace{1em} \widetilde{\lambda}_{1, i}(\infty)  \mspace{5mu} \frac{ \Bigg[-\sqrt{\frac{2}{\pi}} \hspace{0.4em} \rho (1-\mu \beta_{i})+ \sqrt{\frac{2}{\pi} \rho^{2} (1-\mu \beta_{i})^2 + \left(\mu^{2} \beta_{i} \xi^{0} E[\frac{1}{r^{2}}] + \rho^{2}\right) \mu (2-\mu) \beta_{i} } \mspace{4mu} \Bigg] }{ \Bigg[ \sqrt{\frac{2}{\pi} \rho^{2} (1-\mu \beta_{i})^2 + \left(\mu^{2} \beta_{i} \xi^{0} E[\frac{1}{r^{2}}] + \rho^{2}\right) \mu (2-\mu) \beta_{i} } \mspace{4mu} \Bigg] }\\
&= \hspace{1em} \widetilde{\lambda}_{1, i}(\infty) - \widetilde{\lambda}_{1, i}(\infty)\mspace{5mu} \left[\frac{ \sqrt{\frac{2}{\pi}} \hspace{0.4em} \rho (1-\mu \beta_{i}) } { \left[\sqrt{\frac{2}{\pi} \rho^{2} (1-\mu \beta_{i})^2 + \left(\mu^{2} \beta_{i} \xi^{0} E[\frac{1}{r^{2}}] + \rho^{2}\right) \mu (2-\mu) \beta_{i} } \mspace{4mu} \right] }     \right]\\
&=\hspace{1em} \widetilde{\lambda}_{1, i}(\infty) - \widetilde{\lambda}_{1, i}(\infty)\mspace{5mu} \left[\frac{ \sqrt{\frac{2}{\pi}} \hspace{0.4em} \rho  } { \left[\sqrt{\frac{2}{\pi} \rho^{2} + \left(\mu^{2} \beta_{i} \xi^{0} E[\frac{1}{r^{2}}] + \rho^{2}\right) \frac{\mu (2-\mu) \beta_{i}}{(1-\mu \beta_{i})^2} } \mspace{4mu} \right] }     \right]\\
&\simeq \hspace{1em} \widetilde{\lambda}_{1, i}(\infty) - \widetilde{\lambda}_{1, i}(\infty)\mspace{5mu} \left[\frac{ \sqrt{\frac{2}{\pi}} \hspace{0.4em} \rho  } { \sqrt{\frac{\mu^{3} \beta^{2}_{i} (2-\mu) \xi^{0} E[\frac{1}{r^{2}}] }{(1-\mu \beta_{i})^2}} } \mspace{4mu}    \right]    \hspace{4em}  \rho^{2} \ll \rho
\end{split}
\end{equation}
\par
By using $\eqref{eq50}$ and $\eqref{eq53}$, cross EMSE of the proposed combination can be calculated.
\section { Convergence Analysis of $a(n)$ for Various Cases of Sparsity level of the system}
\subsubsection{Non-Sparse Systems}
For non-sparse systems, the set $Z$ ( set of zero taps) contains very small number of coefficients, Therefore from $\eqref{eq35}$ and $\eqref{eq50}$ we have $\widetilde{\lambda}_{1, i}(\infty) \simeq \widetilde{\lambda}_{12, i}(\infty)$ and $\widetilde{\lambda}_{1, i}(\infty) < \widetilde{\lambda}_{2, i}(\infty)$ or equivalently,  $J_{ex, 1}(\infty) \simeq J_{ex, 12}(\infty)$ and $J_{ex, 1}(\infty) < J_{ex, 2}(\infty)$ . These imply $\triangle J_{1}\approx 0$ and $\triangle J_{2}=J_{ex, 2}(\infty)-J_{ex, 12}(\infty)=J_{ex, 2}(\infty)-J_{ex, 1}(\infty) > 0$. Therefore, Eq. $\eqref{eq13}$ leads to
\begin{equation}\label{eq54}
\begin{split}
E[a(n+1)]= E[a(n)] + \mspace{2mu} \mu_{a} \mspace{2mu} E\left[\lambda(n) \mspace{2mu} (1 - \lambda(n))^{2} \right] \Delta J_{2}
\end{split}
\end{equation}
Note that $ \forall \lambda(n) \in [0, 1],$ the function $f(\lambda(n))=\lambda^{2}(n)(1-\lambda(n)) \geq 0$, with a maxima at $\lambda(n)=\frac{2}{3}$ and with $f(\lambda(n))=0$ at $\lambda(n)=0, 1$. Assume that at the $n$-th index, $-a^{+} < E[a(n)] < a^{+}$, meaning that $a(n)$ has not converged to $a^{+}$ or $-a^{+}$ (in all trials), but is taking values from $[-a^{+}, a^{+}]$, or equivalently, $\lambda(n)$ has not converged to $\lambda^{+}$ or $1-\lambda^{+}$ but assuming values from $[(1-\lambda^{+}), \lambda^{+}]$. For $1-\lambda^{+} \leq \lambda(n) \leq \lambda^{+}$, $f(\lambda(n))\geq f(1-\lambda^{+})=C$. Substituting in \eqref{eq54}, $E[a(n+1)]= E[a(n)] + \mspace{2mu} \mu_{a} \mspace{2mu} C \Delta J_{2}$. Since $\triangle J_{2}>0$, the above implies $\lim\limits_{n \to \infty} E[a(n)]=a^{+}$ and thus $\lim\limits_{n \to \infty} a(n)=a^{+}$ almost surely. The results in $\lim\limits_{n \to \infty} \lambda(n)=\lambda^{+}$ (almost surely) $\approx 1$, and therefore the proposed combination switches to APA algorithm (Filter $1$) which performs better than ZA-APA algorithm (Filter $2$) (i.e., $J_{ex, 1} (\infty) < J_{ex, 2} (\infty)$).
\subsubsection{Semi-Sparse Systems}
In semi-sparse systems the zero taps set $Z$ contains significant amount of elements when compared to non-sparse system. For the given controller constant $\rho$ and $0 < \mu < 2$, the term $\left[\frac{ \sqrt{\frac{2}{\pi}} \hspace{0.4em} \rho  } { \sqrt{\frac{\mu^{3} \beta^{2}_{i} (2-\mu) \xi^{0} E[\frac{1}{r^{2}}] }{(1-\mu \beta_{i})^2}} } \mspace{4mu}    \right]$ in the equation $\eqref{eq53}$ becomes positive. Therefore, for every $i \in Z$, we have $\widetilde{\lambda}_{12, i}(\infty) \leq \widetilde{\lambda}_{1, i}(\infty)$ i.e., $J_{ex, 1} (\infty) -J_{ex, 12} (\infty) > 0$. And also, due to the presence of large number of non-zero coefficients still APA performs better than ZA-APA algorithm i.e., $J_{ex, 2} (\infty) > J_{ex, 1} (\infty)$. So, we have $J_{ex, 2} (\infty) > J_{ex, 1} (\infty) > J_{ex, 12} (\infty)$ and thus both $\triangle J_{1} >0$ and $\triangle J_{2} >0$. This is analogous to the case (3), section III of $\cite{18}$. Under this, a stationary point is obtained by setting the update term in \eqref{eq13} to zero as $n \to \infty$, leading to $E[\lambda(\infty)(1-\lambda(\infty))^{2}] \triangle J_{2}=E[\lambda^{2}(\infty)(1-\lambda(\infty))] \triangle J_{1}$. Assuming a negligibly small variance for $\lambda(\infty)$, i.e., assuming $E[\lambda^{2}(\infty)] \to 0 $ which implies that $\lambda(\infty) \to$ a constant (almost surely) as $n \to \infty$ one can then obtain from the above $(1-E[\lambda(\infty)]) \triangle J_{2} = E[\lambda(\infty)] \triangle J_{1}$, or equivalently,
\begin{equation}\label{eq55}
\begin{split}
E[\lambda(\infty)]= \left[\frac{\triangle J_{2}}{\triangle J_{1}+\triangle J_{2}} \right]_{1-\lambda^{+}}^{\lambda^{+}}
\end{split}
\end{equation}
It follows that
\begin{equation}\label{eq56}
\begin{split}
&\lambda^{+} \geq \overline{\lambda}(\infty) > 0.5 ; \mspace{20mu}\textit{if} \mspace{10mu} J_{ex, 1} (\infty) <J_{ex, 2} (\infty)\\
&0.5 \geq \overline{\lambda}(\infty) > 1 - \lambda^{+} ; \mspace{15mu}\mspace{10mu}\textit{if} \mspace{5mu}J_{ex, 1} (\infty) > J_{ex, 2} (\infty)
\end{split}
\end{equation}
As proved in $\cite{18}$, this case is not sub-optimal. Rather it leads to
\begin{equation}\label{eq57}
\begin{split}
J_{ex}(\infty) \leq min[J_{ex, 1}(\infty),J_{ex, 2}(\infty)],
\end{split}
\end{equation}
which means in this case, the proposed convex combination works even better than each of its component filters.
\subsubsection{Sparse Systems}
For sparse systems, Z (set of in active taps) contains the majority of the coefficients and NZ ( set of active taps) contains very negligible number of elements. From $\eqref{eq39}$, we have $\widetilde{\lambda}_{2, i}(\infty) < \widetilde{\lambda}_{1, i}(\infty)$, or equivalently, $J_{ex, 2}(\infty) < J_{ex, 1}(\infty)$, i.e, the ZA-APA algorithm in this case outperforms the APA algorithm. Depending on the value of the controller constant $\rho$, $J_{ex, 2}(\infty) < J_{ex, 12}(\infty)$ and $J_{ex, 2}(\infty) > J_{ex, 12}(\infty)$ are possible. From $\eqref{eq39}$ and $\eqref{eq53}$, we have
\begin{equation}\label{eq58}
\begin{split}
J_{ex, 2}(\infty) \leq J_{ex, 12}(\infty) \mspace{20mu}   \textit{for} \mspace{20mu} 0 < \rho^{2} \leq \frac{ \frac{2}{\pi}\mu^{2}(1-\mu \beta_{i})^{2}\beta_{i}\xi^{0} E[\frac{1}{r^{2}}] }{ \mu (2-\mu)\beta_{i} + \frac{2}{\pi} (1-\mu \beta_{i})^{2} } \\
\end{split}
\end{equation}
Case I is the consequence of the above situation as we discuss in the following paragraph.
\par
Case I : $J_{ex, 2}(\infty) < J_{ex, 12}(\infty) < J_{ex, 1}(\infty) $, that imply $\triangle J_{2} \leq 0$ and $\triangle J_{1}  > 0$\\
This is analogous to the case (1), section III of $\cite{18}$. Equation $\eqref{eq13}$ in this case leads to
\begin{equation}\label{eq59}
\begin{split}
E[a(n+1)]=E[a(n)] + \mu_{a} E[g(\lambda(n))] \triangle J_{2}-\mu_{a} E[f(\lambda(n))] \triangle J_{1},
\end{split}
\end{equation}
where $g(\lambda(n)) = \lambda(n)[1-\lambda(n)]^{2}=f(1-\lambda(n))$. Like $f(\lambda(n))$, $g(\lambda(n)) \geq g(\lambda^{+})= f(1- \lambda^{+})= C, \forall \lambda(n) \in [1-\lambda^{+}, \lambda^{+}]$. From the arguments used for the non-sparse case above, we have, $E[a(n+1)] \leq E[a(n)] - \mu_{a} C (\triangle J_{2}-\triangle J_{1})$, which means $\lim\limits_{n \to \infty} E[a(n)]= - a^{+}$ and thus $\lim\limits_{n \to \infty} a(n)= -a^{+}$ (almost surely), or equivalently, $\lim\limits_{n \to \infty} E[\lambda(n)]= 1-\lambda^{+}$ ( almost surely ) $\approx$ 0. The combination filter in this case will be converged to the ZA-APA based filter which is better of the two filters for sparse system.
\par
For higher values of $\rho$, the cross EMSE eventually becomes relatively less than the EMSE of ZA-APA which gives case II.
\par
Case II : $J_{ex, 12}(\infty) < J_{ex, 2}(\infty) < J_{ex, 2}(\infty) $, meaning $\triangle J_{1} > 0$\\
This is again analogous to the case (3), section III of $\cite{18}$. Using the arguments used for the semi-sparse case above, $\eqref{eq55}$  and thus $\eqref{eq57}$  will be satisfied in this case, meaning the proposed combined filter will perform better than both filter$1$ and filter$2$.
\section{ A New Approach to Increase the Convergence Rate of the Combination in Sparse System Case}
In sparse system case, ZA-APA provides better steady-state EMSE performance, however, cannot improve the convergence rate. The proportionate-type concepts can be incorporated into the ZA-APA algorithm to attain both improved convergence rate and steady-state EMSE performance simultaneously. Therefore, we are using Zero Attracting Proportionate Affine Projection Algorithm (ZA-PAPA) as filter$2$ in the proposed combination. The weight update equation of the ZA-PAPA algorithm is as follows,
\begin{equation}\label{eq60}
\textbf{w}_{2}(n+1)=\textbf{w}_{2}(n)+\mu \hspace{0.5em}\textbf{G}(n) \hspace{0.5em}
{\textbf{U}}(n) \Big(\varepsilon \textbf{I}_{M} + {\textbf{U}}^{T}(n)\textbf{G}(n){\textbf{U}}(n)\Big)^{-1}
{\textbf{e}}_{2}(n) \hspace{0.5em} - \rho \hspace{0.5em} sgn\left( \textbf{w}_{2}(n) \right)
\end{equation}
where $\textbf{G}(n)$ is a diagonal gain
matrix that distributes the adaptation energy unevenly over the
filter taps by modifying the step size of each tap.
\par
The gain matrix $\textbf{G}(n)$ is evaluated as,
\begin{equation} \label{eq61}
\textbf{G}(n)=diag(g_{0}(n),g_{1}(n),...g_{L-1}(n))
\end{equation}
where,
\begin{equation}\label{eq62}
g_{l}(n)=\frac{\gamma_{l}(n)}{\frac{1}{L} \sum\limits_{l=0}^{L-1}\gamma_{l}(n)}, \mspace{8mu} 0\leq l \leq (L-1)
\end{equation}
with,
\begin{eqnarray}\label{eq63}
\begin{split}
\gamma_{ l}(n)=max[\rho\mspace{4mu} \gamma_{min}(n), \mathrm{F}[|w_{l}(n)|]] \\
\gamma_{min}(n)=max(\delta, \mathrm{F}[|w_{0}(n)|],...,\mathrm{F}[|w_{L-1}(n)|]
\end{split}
\end{eqnarray}
where $\rho$ is a very small positive constant which together
with $\gamma_{min}(n)$, ensures that $\gamma_{l}(n)$ and thus
$g_{l}(n)$ do not turn out to be zero for the inactive taps and
thus the corresponding updation does not stall. The parameter
$\delta$ is again a small positive constant employed to avoid
stalling of the weight updation at the start of the iterations
when the tap weights are initialized to zero.
\par
A full understanding of the APA and ZA-PAPA combination requires a steady
state EMSE analysis for $a(n)$ or $\lambda(n)$, which is beyond
the scope of this correspondence. Detailed simulations of the convex
combination of ZA-PAPA and APA are presented hereunder.
\section{Simulation Studies and Discussion}
Here the analytical results presented in section IV are compared with the simulations for system identification example. The proposed algorithm has been simulated for identifying the system ($\textbf{w}_{opt}$) of length $L=256$ taps. Initially the system is taken to be non-sparse with all coefficients (randomly chosen) having significant magnitude. After $6000$ time samples, the system is changed to a semi-sparse system with $80$ active taps and the remaining coefficients magnitude equal to zero. Finally, after $12000$ time samples the system is changed to a highly-sparse system having $16$ active taps with the remaining coefficients being inactive.
\par
Simulations were performed using zero mean, Gaussian white noise and first-order auto-regressive (AR(1)) process having a pole at $0.8$ with unit variance ($\sigma^{2}_{u}=1$ ). The observation noise $\epsilon(n)$ was taken to be zero-mean Gaussian white noise with variance $\xi^{0}=0.001$. Projection order ($M$) was taken to be $8$ for both the APA and the ZA-APA algorithms and the initial taps were chosen to be zero. The controller constant $\rho$ was taken to be $8 \times 10^{-6}$ and $3 \times 10^{-5}$ for white and colored input cases respectively. The simulation results shown in Fig. $2$ and Fig. $3$ are obtained by plotting the EMSE against the iteration index $n$, by averaging over $1000$ experiments.
\par
\begin{figure}[h]
\centering
\includegraphics [height=70mm,width=160mm]{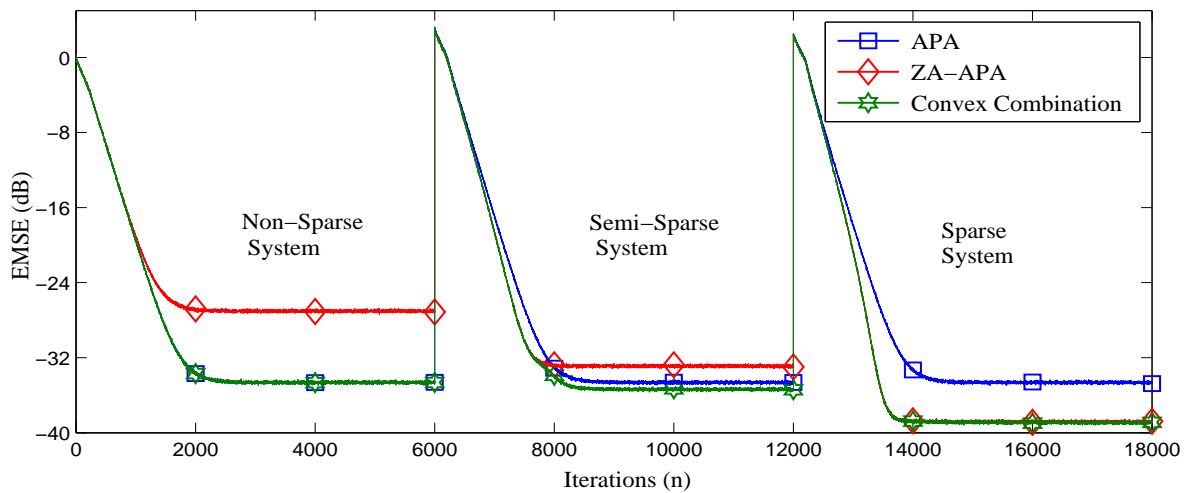}
\caption{Learning curves of APA and ZA-APA combination for white input with $M=8$}
\label{the-label-for-cross-referencing}
\end{figure}
Several interesting observations follow from Fig. $2$ and Fig. $3$. The plots confirm our conjuncture that for non-sparse system $J_{ex, 1}(\infty)\approx J_{ex, 12}(\infty)$ while for all other cases (i.e., sparse and semi-sparse systems), $J_{ex, 1}(\infty) \geq J_{ex, 12}(\infty)$. Secondly, for the chosen value of $\rho$, it is seen that $J_{ex, 2}(\infty) < J_{ex, 12}(\infty)$ when the system is highly-sparse. As per our analysis above, the proposed combination in this case should converge to the ZA-APA based filter, meaning we should have $J_{ex}(\infty) \approx J_{ex, 2}(\infty)$. Similarly, for the non-sparse system, it is seen from Fig. $2$  and Fig. $3$ that $J_{ex}(\infty) \approx J_{ex, 1}(\infty)$, meaning the convex combination in this case favors filter$1$, which confirms our arguments above. Lastly, for the semi-sparse system, the plots confirm that $J_{ex, 2}(\infty) > J_{ex, 1}(\infty)> J_{ex, 12}(\infty)$. As per the discussions of the previous section, the proposed combination in this case is likely to produce a solution that performs better than both filter$1$ and filter$2$.
\begin{figure} [h!]
\centering
\includegraphics [height=70mm,width=160mm]{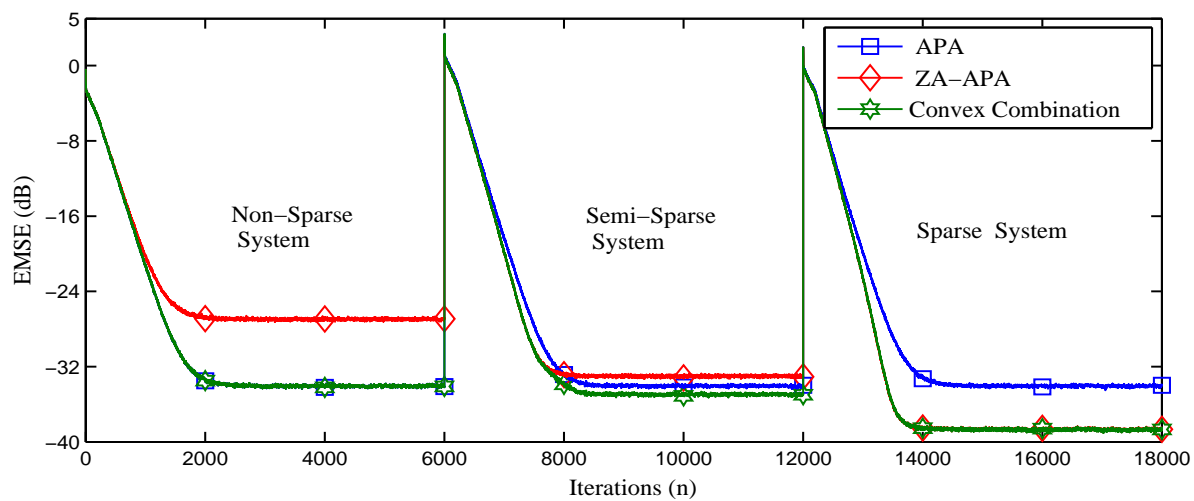}
\caption{Learning curves of APA and ZA-APA combination for AR(1) input having pole $\alpha=0.8$ and $M=8$}
\label{the-label-for-cross-referencing}
\end{figure}
\par
Later, simulations were performed for the proposed convex combination of APA and ZA-PAPA. The learning curves of the proposed combination for white and colored input signal are presented in Fig. $4$ and Fig. $5$ respectively. From these figures, it can be observed that usage of proportionate concepts in ZA-APA provides high convergence rate in sparse system case.
\begin{figure} [h!]
\centering
\includegraphics [height=70mm,width=160mm]{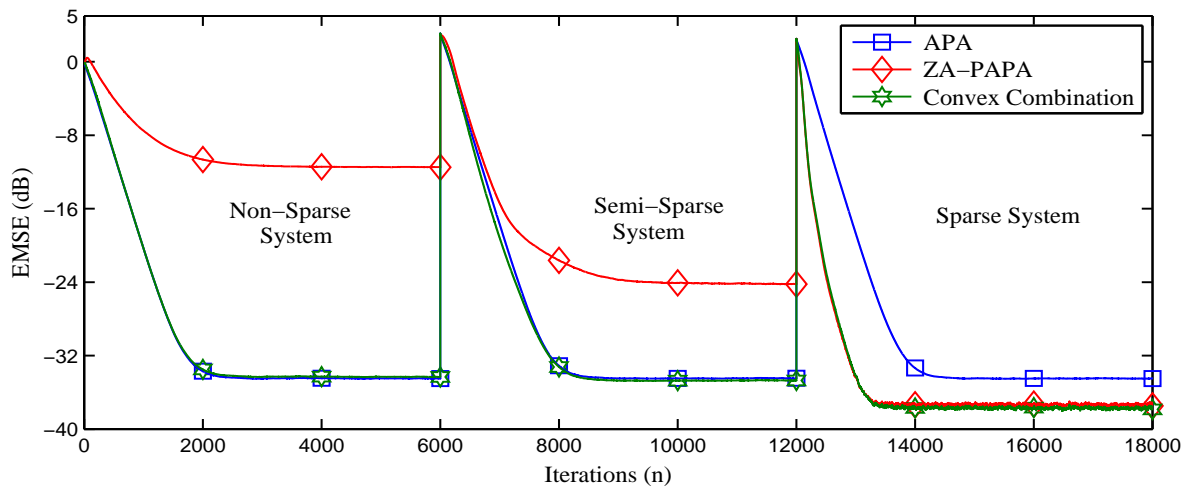}
\caption{Learning curves of ZA-PAPA and APA combination for White input with $M=8$}
\label{the-label-for-cross-referencing}
\end{figure}

\begin{figure} [h!]
\centering
\includegraphics [height=70mm,width=160mm]{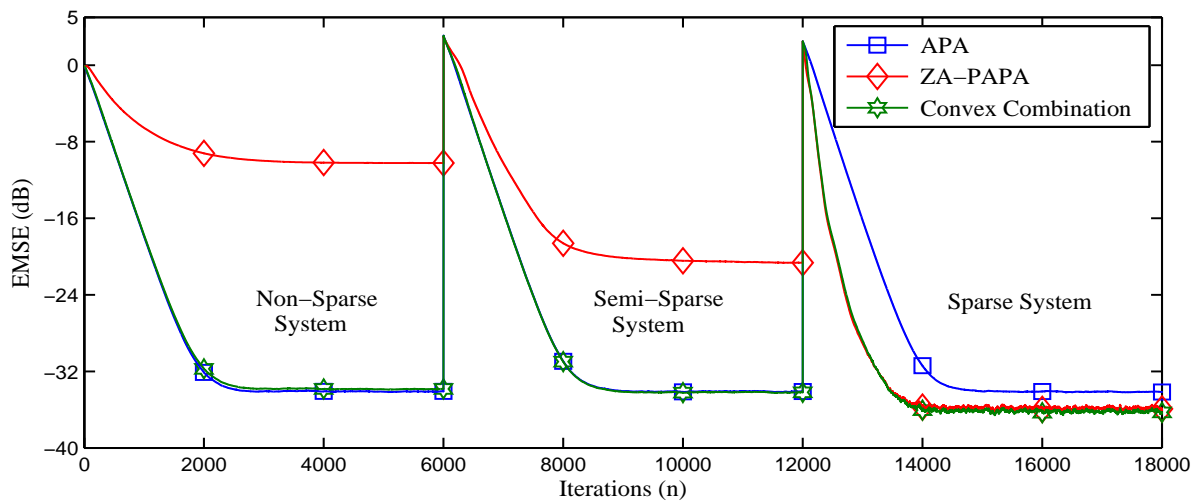}
\caption{Learning curves of ZA-PAPA and APA combination for AR input with $M=8$}
\label{the-label-for-cross-referencing}
\end{figure}
\section{Conclusions}
An algorithm that is robust against the correlated input is presented for identifying sparse systems
with the degree of sparsity varying with time and context .
The proposed method uses an adaptive convex combination
of the "sparsity aware" ZA-APA algorithm
and the standard APA algorithm. The algorithm
adapts dynamically to the level of sparseness and exhibits robustness against the correlated input.

\end{document}